\documentclass[showpacs,amssymb,aps]{revtex4}

\usepackage{graphicx}
\begin{document}

\title{Quantization in a General Light-front Frame}  

\author{Ashok Das$^{a}$ and Silvana Perez$^{b}$}
\affiliation{$^{a}$ Department of Physics and Astronomy,
University of Rochester,
Rochester, NY 14627-0171, USA}
\affiliation{$^{b}$ Departamento de F\'{\i}sica, 
Universidade Federal do Par\'{a}, 
Bel\'{e}m, PA 66075-110, BRAZIL}

\bigskip

\begin{abstract}

In this paper, we study the question of quantization of quantum field
theories in a general light-front frame. We quantize scalar, 
fermion as well as gauge field theories in a systematic manner
carrying out the Hamiltonian analysis carefully. The decomposition of
the fields into positive and negative frequency terms needs to be done
carefully after which we show that the (anti) commutation relations
for the quantum operators become frame independent. The frame
dependence is completely contained in the functions multiplying these
operators in the field decomposition. We derive the propagators from
the vacuum expectation values of the time ordered products of the
fields. 

\end{abstract}


\maketitle

\section{Introduction}

Light-front field theories \cite{dirac} have been studied vigorously in the
past. The conventional light-front frame where one defines
\begin{equation}
 \bar{x}^{\mu} = \left(x^{+}, x^{i}, x^{-}\right),\quad x^{\pm} =
 x^{0} \pm x^{3},\quad i=1,2,\label{conventional}
\end{equation}
also describes the infinite momentum frame where many physical results
simplify considerably \cite{fubini}. The quantization of quantum field
theories on the light-front in the conventional light-front frame
(\ref{conventional}) has been studied in detail in
\cite{kogut,chang,brodsky0,brodsky1}. One of the many advantages of using a
light-front quantization is that a larger number of generators of the
Poincare algebra become kinematical \cite{dirac} leading to a trivial
structure  for the vacuum state \cite{brodsky}. This, in principle,
allows for the possibility of
carrying out nonperturbative studies in a simple manner.

More recently, it has been observed \cite{das0,weldon,das} that 
description of statistical mechanics for
field theories quantized on the light-front prefers an oblique
coordinate frame 
\begin{equation}
\bar{x}^{\mu} = \left(x^{+}, x^{i}, x^{3}\right),\label{oblique}
\end{equation}
with $x^{+}$ defined in (\ref{conventional}). The advantage of this
frame lies in the fact that the associated temperature can be
identified with that for the theory in Minkowski space quantized on
an equal time surface. In general, a frame defined by
\begin{equation}
\bar{x}^{\mu} = \left(\bar{x}^{0}, x^{i}, \bar{x}^{3}\right),\quad
\bar{x}^{0} = x^{0} + x^{3},\quad \bar{x}^{3} = A x^{0} + B
x^{3},\label{general}
\end{equation}
where $A,B$ are real constants \cite{gomis}, does allow for a description of
statistical mechanics as long as $A\pm B\neq 0$. In this case, the
temperature cannot be identified with the temperature
associated with the theory in Minkowski space (quantized on an equal
time surface), although they will be related by a multiplicative
factor. 

Since one of the goals of light-front field theories is to study
nonperturbative phenomena using the simplicity of the vacuum
structure, it is best carried out in the operator
formalism. Similarly, questions such as the zero modes and
spontaneous symmetry breaking play an important role in light-front
field theories and can be systematically studied in an operatorial
formalism. To carry
out such studies at finite temperature one would need to use the
formalism  of
thermo field dynamics where one defines a thermal vacuum starting from
a doubled Hilbert space of the original theory through a Bogoliubov
transformation \cite{umezawa,das1}. It is essential, therefore, that
one understands the
questions of operator quantization to construct the
thermal vacuum in such theories. It is with this goal that we have
chosen to study systematically the quantization of theories in a general
light-front frame in this paper. 

The paper is organized as follows. In section {\bf II}, we give some 
details on the properties of various quantities of interest in the
general light-front frame. In section {\bf III}, we carry out the
classical Hamiltonian analysis for a  scalar field theory in such a
general frame and subsequently quantize this theory. The field
decomposition into positive and negative frequency parts needs to be
done carefully which we discuss. We derive all the necessary relations
and derive 
the Feynman propagator from the vacuum expectation value of the time
ordered product of fields. In section {\bf IV}, we introduce various
properties of the Dirac matrices as well as projection operators in
this general frame and carry out the Hamiltonian analysis. The
Hamiltonian analysis can be carried out both in the full spinor space
or in the projected subspaces and lead to the same results. We only
discuss the analysis in the projected space for simplicity. To
quantize such a theory, we also solve the Dirac equation in this
general frame and obtain the positive and negative energy spinors. The
field decomposition can then be carried out into positive and negative
frequency states much like in the scalar case. We obtain the
quantization conditions and derive the Feynman propagator from the
vacuum expectation value of the time ordered product of fields. In
section {\bf V}, we quantize the non-Abelian gauge field theory in the
light-cone gauge and derive the Feynman propagator. We show that it is
doubly transverse as is the case \cite{brodsky1} in the conventional
light-front quantization using (\ref{conventional}). We conclude
with a brief summary in section {\bf VI}.

\section{General Light-front Frame}

As discussed in (\ref{general}) in the introduction, we define the
general light-front frame as the frame where the coordinates have the
form 
\begin{equation}
\bar{x}^{0} = x^{0}+x^{3},\quad \bar{x}^{3} = A x^{0} + B x^{3},\quad
\bar{x}^{i} = x^{i},\quad i=1,2,\label{coorddef}
\end{equation}
with $A,B$ real constants. Here $\bar{x}^{i}=x^{i}, i=1,2$ are known
as transverse coordinates. For $B=-A=-1$, we have the
conventional light-front frame (\ref{conventional}) whereas for
$A=0,B=1$, we have the
oblique light-front frame (\ref{oblique}) used in the statistical
description of light-front theories \cite{das0,weldon,das}. The new
coordinates in (\ref{coorddef}) are
related to the old Minkowski coordinates through a linear transformation,
\begin{equation}
\bar{x}^{\mu} = L^{\mu}_{\;\;\alpha}\,x^{\alpha},\quad x^{\alpha} =
L^{\alpha}_{\;\;\mu}\, \bar{x}^{\mu},\label{def}
\end{equation}
where
\begin{equation}
L^{\mu}_{\;\;\alpha} = \left(\begin{array}{cccc}
1 & 0 & 0 & 1\\
0 & 1 & 0 & 0\\
0 & 0 & 1 & 0\\
A & 0 & 0 & B
\end{array}\right),\quad L^{\alpha}_{\;\;\mu} = \left(\begin{array}{cccc}
-\frac{B}{A-B} & 0 & 0 & \frac{1}{A-B}\\
0 & 1 & 0 & 0\\
0 & 0 & 1 & 0\\
\frac{A}{A-B} & 0 & 0 & -\frac{1}{A-B}
\end{array}\right).\label{l}
\end{equation}
From definition (\ref{def}), it is easy to see that
\begin{equation}
L^{\mu}_{\;\;\alpha}\,L^{\alpha}_{\;\;\nu} = \delta ^{\mu}_{\;\;\nu},\quad
L^{\alpha}_{\;\;\mu}\, L^{\mu}_{\;\;\beta} =
\delta^{\alpha}_{\;\;\beta},\label{ortho}
\end{equation}
which can also be explicitly checked from the representation in
(\ref{l}). 

Under a change of frame (\ref{def}), it is clear from (\ref{ortho})
that scalars remain invariant while vectors and tensors transform. In
particular, the metric tensor transforms as
\begin{eqnarray}
\bar{g}^{\mu\nu} & = &
L^{\mu}_{\;\;\alpha}\,\eta^{\alpha\beta}\,L^{\nu}_{\,\beta} =
\left(\begin{array}{crrc} 
0 & 0 & 0 & (A-B)\\
0 & -1 & 0 & 0\\
0 & 0 & -1 & 0\\
(A-B) & 0 & 0 & (A^{2}-B^{2})
\end{array}\right),\nonumber\\
\noalign{\vskip 6pt}%
\bar{g}_{\mu\nu} & = &
L_{\;\;\mu}^{\alpha}\,\eta_{\alpha\beta}\,L_{\;\;\nu}^{\beta} =
 = \left(\begin{array}{crrc}
-\frac{A+B}{A-B} & 0 & 0 & \frac{1}{A-B}\\
0 & -1 & 0 & 0\\
0 & 0 & -1 & 0\\ 
\frac{1}{A-B} & 0 & 0 & 0
\end{array}\right),\label{metric}
\end{eqnarray}
where $\eta^{\alpha\beta},\eta_{\alpha\beta}$ represent the usual
Minkowski space metric (with signatures $(+,-,-,-)$). Since the
transformation does not necessarily
represent a Lorentz transformation, the metric transforms in general. It
follows from this that
\begin{equation}
\det (-\bar{g}_{\mu\nu}) = (-\bar{g}) = \frac{1}{(A-B)^{2}} > 0,\quad
\sqrt{-\bar{g}} = \frac{1}{\left|A-B\right|},\quad \sqrt{-\bar{g}}\,
(A-B) = {\rm sgn}\, (A-B).\label{relation}
\end{equation}
We note that for $\bar{x}^{0}$ to represent the time coordinate in the
transformed frame, we must have \cite{weldon}
\begin{equation}
\bar{g}_{00}\geq 0,\quad \Rightarrow \quad |B|\geq
|A|.\label{inequality} 
\end{equation}

Let us next note that a covariant vector transforms under such a
change of frame as
\begin{equation}
\bar{V}_{\mu} = L_{\;\;\mu}^{\alpha}\, V_{\alpha},
\end{equation}
which leads to the transformation of the energy-momentum four-vector as
\begin{equation}
\bar{p}_{0} = \frac{1}{(A-B)}\,\left(-B p_{0} + A p_{3}\right),\quad
\bar{p}_{3} = \frac{1}{(A-B)}\,\left(p_{0}-p_{3}\right),\quad
\bar{p}_{i} = p_{i}.
\end{equation}
The Einstein relation,
\[
\bar{p}^{2} = \bar{g}^{\mu\nu}\bar{p}_{\mu}\bar{p}_{\nu} = m^{2},
\]
in this case, leads to
\begin{eqnarray}
2 (A-B) \bar{p}_{0}\bar{p}_{3} & = & \bar{p}_{i}^{2} + (B^{2}-A^{2})
\bar{p}_{3}^{2} + m^{2}\nonumber\\
\noalign{\vskip 4pt}%
{\rm or,}\quad \bar{p}_{0} & = &
\frac{\omega_{\bar{p}}^{2}}{2(A-B)\bar{p}_{3}},\label{einsteinrelation}
\end{eqnarray}
where we have defined
\begin{equation}
\omega_{\bar{p}}^{2} = \bar{p}_{i}^{2} + (B^{2}-A^{2}) \bar{p}_{3}^{2}
+ m^{2}.\label{omega}
\end{equation}
It is easily seen using (\ref{inequality}) that this quantity is
positive definite and reduces to the corresponding definition in the
conventional light-front frame \cite{kogut,chang} as well as the
oblique light-front frame \cite{das} for particular values of $A,B$
noted earlier.

We note that the invariant volumes in the coordinate
and momentum spaces are given by
\begin{equation}
\int \mathrm{d}^{4}\bar{x}\,\sqrt{-\bar{g}} = \int
\mathrm{d}^{4}x,\quad \int
\mathrm{d}^{4}\bar{p}\,\left(\sqrt{-\bar{g}}\right)^{-1} = \int
\mathrm{d}^{4}p. \label{volume}
\end{equation}
Note also from (\ref{metric}) that
\begin{equation}
\bar{g}_{00} = - \frac{A+B}{A-B},
\end{equation}
which vanishes for the conventional light-front frame
(\ref{conventional}) making a
statistical description impossible while for the oblique light-front
coordinates in (\ref{oblique}) $\bar{g}_{00}=1$ leading to a
statistical description
where the temperature can be identified with that of the original
Minkowski frame \cite{weldon}. We note that for any $A,B$ such that
$A\pm B \neq 0$ 
(namely, $\bar{g}_{00}\neq 0$ or divergent), a statistical description
is possible with a nontrivial scaling of the temperature. In the following
sections, we will quantize scalar, fermion and gauge field theories in
a general light-front frame with arbitrary $A,B$. 

Finally, for completeness, we note that under a Lorentz boost along
the $z$-axis, 
\[
\tilde{x}^{0} = \gamma \left(x^{0} + \beta x^{3}\right),\quad
\tilde{x}^{3} = \gamma \left(x^{3} + \beta x^{0}\right),\quad \gamma =
\cosh \phi,\quad \beta\gamma = \sinh \phi,
\]
so that we have
\begin{equation}
\tilde{\bar{x}}^{0} = e^{\phi}\,\bar{x}^{0},\quad \tilde{\bar{x}}^{3}
= e^{-\phi}\,\bar{x}^{3} + (A+B)\,\sinh \phi\,\bar{x}^{0}.
\end{equation}
We see that such a boost acts as a scale transformation for
$\bar{x}^{0}$, but not for $\bar{x}^{3}$ in general, unless $A+B=0$
corresponding to the conventional light-front frame
(\ref{conventional}). However, the quantization surface
$\bar{x}^{0}=0$ remains invariant under such a
transformation. Similarly, it can be seen that the generators $J_{3},
E_{1} = -K_{1} + J_{2}, E_{2} = -K_{2}-J_{1}$, where $J_{i}$ and
$K_{i}$ correspond to angular momentum and boost operators
respectively, also leave the surface
of quantization invariant. Consequently, $J_{3}, K_{3}, E_{1}, E_{2}$
correspond to kinematical generators much like in the conventional
light-front frame \cite{harindranath}. In addition, it is clear that
translations along the $x^{i}, (x^{0}-x^{3})$ leave the quantization
surface invariant leading to the fact that $P_{i}, (P_{0}-P_{3})$ are
kinematical generators as well \cite{brodsky}.

\section{Scalar Fields}

In this section, we will carry out the Hamiltonian analysis for the
scalar field and quantize it in the general light-front frame. Using
the transformation laws discussed in the previous section, it is
easily seen that the action for a free scalar field can be written as
\begin{equation}
S = \int \mathrm{d}^{4}\bar{x}\,\sqrt{-\bar{g}}\,{\cal L},
\end{equation}
where
\begin{eqnarray}
{\cal L} & = & \frac{1}{2}\left(\bar{g}^{\mu\nu}
\bar{\partial}_{\mu}\phi\,\bar{\partial}_{\nu}\phi -
m^{2}\phi^{2}\right)\nonumber\\
\noalign{\vskip 4pt}%
& = & (A-B) \bar{\partial}_{0}\phi \bar{\partial}_{3}\phi -
\frac{1}{2} \left(\bar{\partial}_{i}\phi\right)^{2} - \frac{1}{2}
(B^{2}-A^{2}) \left(\bar{\partial}_{3}\phi\right)^{2} -
\frac{m^{2}}{2}\,\phi^{2}.\label{action1}
\end{eqnarray}
The conjugate momentum density can now be defined in the standard
manner as
\begin{equation}
\Pi = \frac{\partial \left(\sqrt{-\bar{g}}\, {\cal L}\right)}{\partial
\bar{\partial}_{0}\phi} = \sqrt{-\bar{g}}\, (A-B) \bar{\partial}_{3}\phi
= {\rm sgn}\, (A-B)\,\bar{\partial}_{3}\phi,
\end{equation}
where we have used (\ref{relation}). This leads to the only primary
constraint of the theory of the form \cite{dirac} 
\begin{equation}
\chi = \Pi - {\rm sgn}\, (A-B)\,\bar{\partial}_{3}\phi \approx
0.\label{primary1}
\end{equation}

The Hamiltonian density together with the primary constraint takes the
form 
\begin{equation}
{\cal H} =
\frac{\sqrt{-\bar{g}}}{2}\left(\left(\bar{\partial}_{i}\phi\right)^{2}
+ (B^{2}-A^{2}) \left(\bar{\partial}_{3}\phi\right)^{2} +
m^{2}\phi^{2}\right) + \lambda \chi,
\end{equation}
where $\lambda$ represents the Lagrange multiplier to be
determined. Requiring the primary constraint to be stationary, we
obtain
\begin{eqnarray}
\dot{\chi} (\bar{x}) & = & \left\{\chi (\bar{x}), H\right\}\nonumber\\
\noalign{\vskip 4pt}%
 & = & \sqrt{-\bar{g}}\left(\bar{\partial}_{i}^{2}\phi + (B^{2}-A^{2})
\bar{\partial}_{3}^{2}\phi - m^{2}\phi - 2(A-B)
\bar{\partial}_{3}\phi\right) \approx 0,\label{stationary}
\end{eqnarray}
where a ``dot'' denotes derivative with respect to $\bar{x}^{0}$ and
in evaluating the Poisson bracket above, we have used the canonical
Poisson brackets between variables (for equal $\bar{x}^{0}$
coordinates), namely,
\begin{equation}
\left\{\phi (\bar{x}), \phi (\bar{y})\right\} = 0 = \left\{\Pi
(\bar{x}), \Pi (\bar{y})\right\},\quad \left\{\phi (\bar{x}), \Pi
(\bar{y})\right\} = \delta^{3} (\bar{x}-\bar{y}).
\end{equation}

Equation (\ref{stationary}) determines the Lagrange multiplier
$\lambda$ and shows that there are no further
constraints in the theory except
for the primary constraint. The equal ``time'' ($\bar{x}^{0}$) Poisson
bracket between the primary constraint leads to the matrix
\begin{equation}
\left\{\chi (\bar{x}), \chi (\bar{y})\right\} = C(\bar{x},\bar{y}) =
-2 {\rm sgn}\, (A-B) \bar{\partial}_{3}^{(\bar{x})} \delta^{2}
(\bar{x}_{\perp}-\bar{y}_{\perp}) \delta (\bar{x}^{3}-\bar{y}^{3}),
\end{equation}
where we have identified collectively $\bar{x}_{\perp} = (\bar{x}^{i})$. The
inverse of this matrix of constraints is easily obtained to be
\begin{equation}
C^{-1} (\bar{x},\bar{y}) = - \frac{{\rm sgn}\, (A-B)}{2}\,\delta^{2}
(\bar{x}_{\perp} - \bar{y}_{\perp}) \epsilon
(\bar{x}^{3}-\bar{y}^{3}),
\end{equation}
and we have defined the alternating step function as 
\begin{equation}
\epsilon (x) = \frac{1}{2} \left(\theta (x) - \theta
(-x)\right),
\end{equation}
such that
\begin{equation}
\partial \epsilon (x) = \delta (x).
\end{equation}

With this the Dirac brackets between the variables can be
calculated. The independent (equal time) bracket takes the form
\begin{equation}
\left\{\phi (\bar{x}), \phi(\bar{y})\right\}_{\rm D} = - \frac{{\rm sgn}\,
(A-B)}{2}\, \delta^{2} (\bar{x}_{\perp} - \bar{y}_{\perp}) \epsilon
(\bar{x}^{3} - \bar{y}^{3}).\label{dirac1}
\end{equation}
Since the constraints can be set strongly equal to zero in the Dirac
brackets, the brackets between other variables can be easily obtained
from this using the primary constraint (\ref{primary1}). Furthermore,
in going to the quantum theory, we can obtain the basic, independent
(equal time) commutation relation between fields from (\ref{dirac1})
to be 
\begin{equation}
\left[\phi (\bar{x}), \phi (\bar{y})\right] = - \frac{i\,{\rm sgn}\,
(A-B)}{2}\, \delta^{2} (\bar{x}_{\perp} - \bar{y}_{\perp}) \epsilon
(\bar{x}^{3}-\bar{y}^{3}),\label{comm1}
\end{equation}
where we have assumed $\hbar =1$ (which we will assume throughout this
paper).

In the general light-front frame, the decomposition of the fields into
positive and negative energy parts has to be done carefully and since
it is essential for the subsequent discussions, we discuss this in
some detail. We note that since the scalar field satisfies the equation
\[
\left(\bar{g}^{\mu\nu}\bar{\partial}_{\mu}\bar{\partial}_{\nu} +
m^{2}\right) \phi (\bar{x}) = 0,
\]
the field decomposition takes the form
\begin{eqnarray}
\phi (\bar{x}) & = & \frac{1}{(2\pi)^{3/2}} \int
\mathrm{d}^{4}\bar{k}\,\left(\sqrt{-g}\right)^{-1}\,\delta
(\bar{k}^{2}-m^{2})\,e^{-i\bar{k}\cdot \bar{x}}\,\hat{\phi}
(\bar{k})\nonumber\\
\noalign{\vskip 4pt}%
 & = & \frac{1}{(2\pi)^{3/2}} \int
\mathrm{d}^{4}\bar{k}\,\left(\sqrt{-\bar{g}}\right)^{-1}\, 
\delta \left(2(A-B)\bar{k}_{0}\bar{k}_{3} -
\omega_{\bar{k}}^{2}\right)\,e^{-i\bar{k}\cdot
\bar{x}}\,\hat{\phi} (\bar{k}),\label{decomp}
\end{eqnarray}
where $\omega_{\bar{k}}$ is defined in (\ref{omega}). We note that the
delta function constraint leads to
\[
\bar{k}_{0} = \frac{\omega_{\bar{k}}^{2}}{2(A-B)\bar{k}_{3}}.
\]
As a result, the sign of the energy depends on the sign of $(A-B)$ and
if we integrate out $\bar{k}_{0}$, we cannot obtain a clean separation
into 
positive and negative energy terms in the usual manner. Let us,
therefore, scale and define
\begin{equation}
\left(\bar{k}_{i}, \bar{k}_{3}\right)\rightarrow {\rm sgn}\,(A-B)
\left(\bar{k}_{i}, \bar{k}_{3}\right),\quad \tilde{\bar{k}} =
\left(\bar{k}_{0}, {\rm sgn}\,(A-B) \bar{k}_{i}, {\rm sgn}\,(A-B)
\bar{k}_{3}\right),\label{ktilde}
\end{equation}
so that the field decomposition (\ref{decomp}) can be written as
\begin{eqnarray}
\phi (\bar{x}) & = & \frac{1}{(2\pi)^{3/2}} \int
\mathrm{d}^{2}\bar{k}_{\perp} \int_{-\infty}^{\infty}
\mathrm{d}\bar{k}_{3}\,\frac{\left(\sqrt{-\bar{g}}\right)^{-1}}
{2|A-B||\bar{k}_{3}|}\,e^{-i\tilde{\bar{k}}\cdot
  \bar{x}}\,\hat{\phi} (\tilde{\bar{k}})\nonumber\\
\noalign{\vskip 4pt}%
 & = & \frac{1}{(2\pi)^{3/2}} \int \mathrm{d}^{2}\bar{k}_{\perp}
\int_{0}^{\infty}
\frac{\mathrm{d}\bar{k}_{3}}{2\bar{k}_{3}}\,\left(e^{-i\tilde{\bar{k}}\cdot
  \bar{x}}\, a(\bar{k}) + e^{i\tilde{\bar{k}}\cdot
 \bar{x}}\,a^{\dagger} (\bar{k})\right),\label{decom1}
\end{eqnarray}
where we have used (\ref{relation}) and have identified
\begin{equation}
\bar{k}_{0} = \bar{\omega}=
\frac{\omega_{\bar{k}}^{2}}{2|A-B|\bar{k}_{3}} >0,\quad
a(\bar{k}) = \hat{\phi} (\tilde{\bar{k}}),\quad a^{\dagger}
(\bar{k}) = \hat{\phi} (-\tilde{\bar{k}}).\label{energy}
\end{equation}
With this, the decomposition of the field into positive and negative
energy parts
is now complete and, consequently, we can identify $a(\bar{k})$ and
$a^{\dagger} (\bar{k})$ with annihilation and creation operators
respectively. 
 
Requiring the field variables to satisfy the commutation relation 
(\ref{comm1}), it can be easily determined that the basic non-vanishing
commutation relation of the operators $a,a^{\dagger}$ takes the form
\begin{equation}
\left[a (\bar{k}), a^{\dagger} (\bar{k}')\right] = 2\,\bar{k}_{3}
\delta^{3} (\bar{k}-\bar{k}').\label{basiccomm1}
\end{equation}
This shows that the basic commutation relation between the operators
$a(\bar{k}), a^{\dagger} (\bar{k})$ remains the same in any general
light-front frame (when the decomposition into positive and negative
energy parts has been properly carried out) and the frame dependence is
really contained in the spatial part of the plane wave solutions. We
would like to emphasize here that had we not carefully carried out the
decomposition into positive and negative frequency parts through the
use of (\ref{ktilde}) and (\ref{energy}), the commutation relation
(\ref{basiccomm1}) would involve a factor of ${\rm sgn}\,(A-B)$. For
bosonic theories, this is not a problem and would simply imply that
depending on the sign of $(A-B)$, the roles of $a,a^{\dagger}$ have to
be interchanged. However, the problem is more serious for theories
involving fermions where one cannot use an expansion involving
positive and negative spinors until a careful separation into positive
and negative energy parts has been carried out.

Given the basic commutation relations (\ref{basiccomm1}), the
two point function can now be calculated easily and leads to
\begin{equation}
\langle 0|\phi(\bar{x}) \phi (\bar{y})|0\rangle = \frac{1}{(2\pi)^{3}}
\int \mathrm{d}^{2}\bar{k}_{\perp} \int_{0}^{\infty}
\frac{\mathrm{d}\bar{k}_{3}}{2\bar{k}_{3}}\,e^{-i\tilde{\bar{k}}\cdot
(\bar{x}-\bar{y})}.
\end{equation}
Using the integral representation for the step function
\begin{equation}
\theta (\bar{x}^{0}) = \frac{1}{2\pi i} \int_{-\infty}^{\infty}
\mathrm{d}\omega'\,\frac{e^{i\omega' \bar{x}^{0}}}{\omega' -
i\epsilon},\label{stepfunction}
\end{equation}
it can be shown with some algebra that the Feynman Green's function of 
the theory has the form
\begin{eqnarray}
iG_{F} (\bar{x}-\bar{y}) & = & \langle 0|T\left(\phi (\bar{x}) \phi
(\bar{y})\right)|0\rangle = \theta (\bar{x}^{0}-\bar{y}^{0}) \langle
0|\phi(\bar{x})\phi (\bar{y})|0\rangle + \theta
(\bar{y}^{0}-\bar{x}^{0}) \langle 0|\phi (\bar{y}) \phi
(\bar{x})|0\rangle\nonumber\\
\noalign{\vskip 4pt}%
 & = & \int
\frac{\mathrm{d}^{4}\bar{k}\,\left(\sqrt{-g}\right)^{-1}}{(2\pi)^{4}}\,
\frac{i}{\bar{k}^{2}-m^{2}+i\epsilon}\,e^{-i\bar{k}\cdot
(\bar{x}-\bar{y})}.
\end{eqnarray}
Consequently, we can identify the momentum space Feynman propagator as
\begin{equation}
iG_{F} (\bar{k}) = \frac{i}{\bar{k}^{2}-m^{2}+i\epsilon} =
\frac{i}{2(A-B)\bar{k}_{0}\bar{k}_{3} - \omega_{\bar{k}}^{2} +
i\epsilon}.
\end{equation}
This coincides with the conventional light-front propagator
\cite{chang}  when
$B=-A=-1$ as well as with the propagator in the oblique coordinates
\cite{das0,das} when $A=0,B=1$.

\section{Fermion Fields}

In dealing with fermion theories, we note that the Dirac matrices,
$\gamma^{\mu}$, would transform like coordinate vectors \cite{das} so
that in the new frame we have
\begin{equation}
\bar{\gamma}^{\mu} = L^{\mu}_{\;\;\alpha}\,\gamma^{\alpha}.
\end{equation}
Explicitly, this leads to 
\begin{equation}
\bar{\gamma}^{0} = \gamma^{0} + \gamma^{3},\quad \bar{\gamma}^{3} = A
\gamma^{0} + B \gamma^{3},\quad \bar{\gamma}^{i} =
\gamma^{i}.\label{gamma} 
\end{equation}
The transformed matrices satisfy the Clifford algebra
\begin{equation}
\left\{\bar{\gamma}^{\mu}, \bar{\gamma}^{\nu}\right\} = 2
\bar{g}^{\mu\nu},
\end{equation}
and from the form of the metric in (\ref{metric}), this leads
to
\begin{eqnarray}
\left(\bar{\gamma}^{0}\right)^{2} & = & 0,\nonumber\\
\noalign{\vskip 4pt}%
\left\{\bar{\gamma}^{0},\bar{\gamma}^{3}\right\} & = & 2
(A-B),\nonumber\\
\noalign{\vskip 4pt}%
\left(\bar{\gamma}^{3}\right)^{2} & = & A^{2} - B^{2} < 0,\nonumber\\
\noalign{\vskip 4pt}%
\left\{\bar{\gamma}^{i},\bar{\gamma}^{j}\right\} & = & 2
\eta^{ij},\qquad i,j=1,2,\nonumber\\
\noalign{\vskip 4pt}%
\left\{\bar{\gamma}^{i},\bar{\gamma}^{0}\right\} & = & 0 =
\left\{\bar{\gamma}^{i}, \bar{\gamma}^{3}\right\},\label{gammatransf}
\end{eqnarray}
where we have used (\ref{inequality}).

Given the transformed Dirac matrices, let us define two projection
operators,
\begin{equation}
P^{+} = \frac{1}{2(A-B)}\,\bar{\gamma}^{3}\bar{\gamma}^{0},\quad P^{-}
= \frac{1}{2(A-B)}\,
\bar{\gamma}^{0}\bar{\gamma}^{3}.\label{projection}
\end{equation}
It is easy to check that these satisfy
\begin{equation}
\left(P^{+}\right)^{2} = P^{+},\quad \left(P^{-}\right)^{2} =
P^{-},\quad P^{\pm}P^{\mp} = 0,\quad P^{+} + P^{-} = \mathbf{1}.
\end{equation}
Thus, these define orthogonal projection operators for any value of
the constants $A,B$. The transformed Dirac matrices do not have very
simple Hermiticity properties. For example,
\begin{eqnarray}
\left(\bar{\gamma}^{0}\right)^{\dagger} & = &
\frac{1}{A-B}\left(-(A+B) \bar{\gamma}^{0} + 2
\bar{\gamma}^{3}\right),\nonumber\\
\noalign{\vskip 4pt}%
\left(\bar{\gamma}^{3}\right)^{\dagger} & = & \frac{1}{A-B}\left(-2AB
\bar{\gamma}^{0} + (A+B) \bar{\gamma}^{3}\right),\nonumber\\
\noalign{\vskip 4pt}%
\left(\bar{\gamma}^{i}\right)^{\dagger} & = & -
\bar{\gamma}^{i}.\label{hermiticity}
\end{eqnarray}
In spite of this, the projection operators can be easily checked to be
Hermitian, namely,
\begin{equation}
\left(P^{+}\right)^{\dagger} = P^{+},\quad
\left(P^{-}\right)^{\dagger} = P^{-}.
\end{equation}
The projection operators can also be seen to satisfy various useful
relations,
\begin{equation}
P^{+} \bar{\gamma}^{0} = 0 = \bar{\gamma}^{0} P^{-},\quad
\bar{\gamma}^{3} P^{+} = P^{-}\bar{\gamma}^{3},\quad \bar{\gamma}^{0}
P^{+} = P^{-}\bar{\gamma}^{0},\quad \bar{\gamma}^{3}P^{-} =
P^{+}\bar{\gamma}^{3},\quad \bar{\gamma}^{i}P^{\pm} =
P^{\pm}\bar{\gamma}^{i},\label{projectionrelation}
\end{equation}
and so on. 

For completeness, we note here that if we use the Bjorken-Drell
representation \cite{bjorken} for the original Dirac matrices, then
the transformed
ones will have the explicit forms (we do not write the form of
$\bar{\gamma}^{i}$ which remains the same)
\begin{eqnarray}
\bar{\gamma}^{0} & = & \left(\begin{array}{cc}
\mathbf{1} & \sigma_{3}\\
\noalign{\vskip 4pt}%
-\sigma_{3} & - \mathbf{1}
\end{array}\right),\nonumber\\
\noalign{\vskip 4pt}%
\bar{\gamma}^{3} & = & \left(\begin{array}{cc}
A\mathbf{1} & B\sigma_{3}\\
\noalign{\vskip 4pt}%
-B\sigma_{3} & - A\mathbf{1}
\end{array}\right),\nonumber\\
\noalign{\vskip 4pt}%
P^{\pm} & = & \frac{1}{2}\left(\begin{array}{cc}
\mathbf{1} & \pm \sigma_{3}\\
\noalign{\vskip 4pt}%
\pm \sigma_{3} & \mathbf{1}
\end{array}\right).\label{representation}
\end{eqnarray}
Here each of the elements represents a $2\times 2$ matrix. The
important thing to note here is that the projection operators,
$P^{\pm}$, are independent of the values of the constants $A,B$.

With these basics, we note that the action for a free, massive fermion
can be written as
\begin{eqnarray}
S & = & \int \mathrm{d}^{4}x\,{\cal L} = \int
\mathrm{d}^{4}x\,\overline{\psi}\left(i\gamma^{\alpha}\partial_{\alpha}
- m\right)\psi\nonumber\\
\noalign{\vskip 4pt}%
 & = &  \int
\mathrm{d}^{4}\bar{x}\,\sqrt{-\bar{g}}\,\psi^{\dagger}\left[2i
P^{+}\bar{\partial}_{0} + i\left((A+B)P^{+} +
(A-B)P^{-}\right)\bar{\partial}_{3} + i\alpha_{i}\bar{\partial}_{i} -
\frac{m}{A-B} \left(-B\bar{\gamma}^{0} +
\bar{\gamma}^{3}\right)\right]\psi,\label{diracaction}
\end{eqnarray}
where we have defined as usual
\begin{equation}
\alpha_{i} = \gamma^{0}\gamma^{i}.
\end{equation}
Let us define the projected fermions
\begin{equation}
\psi_{\pm} = P^{\pm} \psi.\label{fermionprojections}
\end{equation}
The Hamiltonian analysis for the Dirac theory can be carried out
either in terms of the projected
spinor fields or in terms of the original spinor field and they lead
to 
the same result. (We have carried out both of these analyses.) However,
for simplicity,  we will only describe here the
Hamiltonian analysis in terms of the projected spinor fields.

In terms of the projected spinor fields, the action for the free
fermion field (\ref{diracaction}) can be written as
\begin{eqnarray}
S  & = & \int
\mathrm{d}^{4}\bar{x}\,\sqrt{-\bar{g}}\left[\psi_{+}^{\dagger}
\left(2i\bar{\partial}_{0} + i (A+B)\bar{\partial}_{3}\right)\psi_{+}
+ i (A-B) \psi_{-}^{\dagger}
\bar{\partial}_{3}\psi_{-}\right.\nonumber\\
\noalign{\vskip 4pt}%
& & \qquad \left. +
i\psi_{+}^{\dagger}\alpha_{i}\bar{\partial}_{i}\psi_{-} +
i\psi_{-}^{\dagger}\alpha_{i}\bar{\partial}_{i}\psi_{+} -
\frac{m}{A-B} \psi_{+}^{\dagger}\bar{\gamma}^{3}\psi_{-} - \frac{m}{2}
\psi_{-}^{\dagger}
\bar{\gamma}^{0}\psi_{+}\right].\label{diracaction1}
\end{eqnarray}
Using (\ref{hermiticity}) it can be checked that the action is Hermitian.

The canonical momentum densities can now be determined from the action
and lead to (we use a left derivative for the fermions)
\begin{eqnarray}
\Pi_{+}^{\dagger} & = & \frac{\partial \left(\sqrt{-\bar{g}}{\cal
L}\right)}{\partial \bar{\partial}_{0}\psi_{+}} =
-2i\sqrt{-\bar{g}}\,\psi_{+}^{\dagger},\nonumber\\
\noalign{\vskip 4pt}%
\Pi_{+} & = & \frac{\partial \left(\sqrt{-\bar{g}}{\cal
L}\right)}{\partial \bar{\partial}_{0}\psi_{+}^{\dagger}} =
0,\nonumber\\ 
\noalign{\vskip 4pt}%
\Pi_{-}^{\dagger} & = & \frac{\partial \left(\sqrt{-\bar{g}}{\cal
L}\right)}{\partial \bar{\partial}_{0}\psi_{-}} = 0,\nonumber\\
\noalign{\vskip 4pt}%
\Pi_{-} & = & \frac{\partial \left(\sqrt{-\bar{g}}{\cal
L}\right)}{\partial \bar{\partial}_{0}\psi_{-}^{\dagger}} =
0,\label{momenta2}
\end{eqnarray}
where we have suppressed the spinor indices for simplicity. We note
that there are four primary constraints in the theory, namely,
\begin{equation}
\chi_{1}^{\dagger} = \Pi_{+}^{\dagger} +
2i\sqrt{-\bar{g}}\,\psi_{+}^{\dagger}\approx 0,\quad \chi_{2} =
\Pi_{+}\approx 0,\quad \chi_{3}^{\dagger} = \Pi_{-}^{\dagger}\approx
0,\quad \chi_{4} = \Pi_{-}\approx 0.\label{primary2}
\end{equation}
Consequently, adding the primary constraints, we can write the
Hamiltonian density for the theory to be
\begin{eqnarray}
{\cal H} & = & \sqrt{-\bar{g}}\Big[-i(A+B)
\psi_{+}^{\dagger}\bar{\partial}_{3}\psi_{+} - i(A-B)
\psi_{-}^{\dagger}\bar{\partial}_{3}\psi_{-} -
i\psi_{+}^{\dagger}\alpha_{i}\bar{\partial}_{i}\psi_{-} - i
\psi_{-}^{\dagger}\alpha_{i}\bar{\partial}_{i}\psi_{+}\nonumber\\
\noalign{\vskip 4pt}%
 &  & \qquad + \frac{m}{A-B} \psi_{+}^{\dagger}\bar{\gamma}^{3}\psi_{-}
+ \frac{m}{2}\psi_{-}^{\dagger}\bar{\gamma}^{0}\psi_{+}\Big] +
\chi_{1}^{\dagger}\lambda_{1} + \lambda_{2}^{\dagger}\chi_{2} +
\chi_{3}^{\dagger}\lambda_{3} + \lambda_{4}^{\dagger}\chi_{4}.
\end{eqnarray}
Here
$\lambda_{1},\lambda_{2}^{\dagger},\lambda_{3},\lambda_{4}^{\dagger}$
represent the Lagrange multipliers with obvious projections.

The Hamiltonian analysis can now be carried out using the canonical
equal time (equal $\bar{x}^{0}=\bar{y}^{0}$) Poisson brackets
\begin{equation}
\left\{\psi_{\pm, a} (\bar{x}), \Pi_{\pm, b}^{\dagger}
(\bar{y})\right\} = - \left(P^{\pm}\right)_{ab}\,\delta^{3}
(\bar{x}-\bar{y}),\quad \left\{\psi_{\pm, a}^{\dagger} (\bar{x}),
\Pi_{\pm, b} (\bar{y})\right\} = -
\left(P^{\pm}\right)_{ba}\,\delta^{3} (\bar{x}-\bar{y}),\quad
a,b=1,2,3,4,
\end{equation}
with all others vanishing. Requiring the primary constraints
(\ref{primary2}) to be stationary determines the Lagrange multipliers
$\lambda_{1},\lambda_{2}^{\dagger}$ and leads to two secondary
constraints 
\begin{equation}
\chi_{5}^{\dagger} = i(A-B) \bar{\partial}_{3}\psi_{-}^{\dagger} + i
\bar{\partial}_{i} \psi_{+}^{\dagger}\alpha_{i} + \frac{m}{A-B}
\psi_{+}^{\dagger} \bar{\gamma}^{3} \approx 0,\quad \chi_{6} = i
(A-B)\bar{\partial}_{3}\psi_{-} + i
\alpha_{i}\bar{\partial}_{i}\psi_{+} - \frac{m}{2}
\bar{\gamma}^{0}\psi_{+} \approx 0.\label{secondary2}
\end{equation}
Requiring these to be stationary determines the remaining two Lagrange
multipliers $\lambda_{3},\lambda_{4}^{\dagger}$ and the chain of
constraints terminates. Thus, there are six constraints
(\ref{primary2}) and (\ref{secondary2}) and it can be easily verified
that they are all second class. The Dirac brackets can now be obtained
iteratively in a systematic manner and we note the final form of the
nontrivial (equal time) Dirac brackets involving the field variables,
\begin{eqnarray}
\left\{\psi_{+, a} (\bar{x}), \psi_{+, b}^{\dagger} (\bar{x})\right\}_{D}
& = & - \frac{i}{2\sqrt{-\bar{g}}}\,
\left(P^{+}\right)_{ab}\,\delta^{3} (\bar{x}-\bar{y}),\nonumber\\
\noalign{\vskip 4pt}%
\left\{\psi_{-, a} (\bar{x}), \psi_{+, b}^{\dagger} (\bar{y})\right\}_{D}
& = & \frac{{\rm sgn}\,
(A-B)}{2}\,\left(P^{-}\left(i\alpha_{i}\bar{\partial}_{i} -
\frac{m}{2}\bar{\gamma}^{0}\right)P^{+}\right)_{ab}\,\delta^{2}
(\bar{x}_{\perp}-\bar{y}_{\perp}) \epsilon
(\bar{x}^{3}-\bar{y}^{3}),\nonumber\\
\noalign{\vskip 4pt}%
\left\{\psi_{+, a} (\bar{x}), \psi_{-, b}^{\dagger} (\bar{y})\right\}_{D}
& = & \frac{{\rm sgn}\,
(A-B)}{2}\,\left(P^{+}\left(i\alpha_{i}\bar{\partial}_{i} -
\frac{m}{A-B}\bar{\gamma}^{3}\right)P^{-}\right)_{ab}\,\delta^{2}
(\bar{x}_{\perp}-\bar{y}_{\perp}) \epsilon
(\bar{x}^{3}-\bar{y}^{3}),\nonumber\\
\noalign{\vskip 4pt}%
\left\{\psi_{-, a} (\bar{x}), \psi_{-, b}^{\dagger} (\bar{y})\right\}_{D}
& = & \frac{i}
{2|A-B|}\,\left(P^{-}\left(i\alpha_{i}\bar{\partial}_{i} -
\frac{m}{2}\bar{\gamma}^{0}\right)\left(i\alpha_{j}\bar{\partial}_{j}
- \frac{m}{A-B} \bar{\gamma}^{3}\right)P^{-}\right)_{ab}\,\delta^{2}
(\bar{x}_{\perp}-\bar{y}_{\perp})\nonumber\\
\noalign{\vskip 4pt}%
 &  & \quad \times \int \mathrm{d}\bar{z}^{3}\,\epsilon
(\bar{x}^{3}-\bar{z}^{3}) \epsilon(\bar{z}^{3}-\bar{y}^{3}).
\end{eqnarray}
For $B=-A=-1$, these can be seen to coincide with the Dirac brackets
\cite{srivastava} 
derived in the conventional light-front frame
(\ref{conventional}). Since the constraints can now be set equal to
zero  strongly and
$\psi_{-}$ is a constrained field (see (\ref{secondary2})), only the
first of these relations is independent. Every other bracket can be
derived from this using the constraint relations. In going over to the
quantum theory, we can take over the Dirac brackets to
anti-commutation relations and the nontrivial equal time
anti-commutation relation takes the form
\begin{equation}
\left[\psi_{+, a} (\bar{x}), \psi_{+, b}^{\dagger} (\bar{y})\right]_{+} =
\frac{1}{2\sqrt{-\bar{g}}}\, \left(P^{+}\right)_{ab}\,\delta^{3}
(\bar{x}-\bar{y}).\label{comm2}
\end{equation}

The spinor solutions for the theory can be worked out in the projected
space quite
easily (we have worked these out in the full theory as well and they
are completely equivalent). We note that in the plane wave basis, the
Dirac equation takes the form
\begin{eqnarray}
\left(2\bar{k}_{0} + (A+B)\bar{k}_{3}\right) u_{+} (\bar{k}) +
\left(\alpha_{i}\bar{k}_{i} - \frac{m}{A-B} \bar{\gamma}^{3}\right)
u_{-} (\bar{k}) & = & 0,\nonumber\\
\noalign{\vskip 4pt}%
(A-B)\bar{k}_{3} u_{-} (\bar{k}) + \left(\alpha_{i}\bar{k}_{i} -
\frac{m}{2} \bar{\gamma}^{0}\right) u_{+} (\bar{k}) & = &
0.\label{equation2} 
\end{eqnarray}
The spinors can be easily checked to satisfy the Einstein relation
(\ref{einsteinrelation}). 
Here by definition, the projected spinors $u_{\pm} = P^{\pm} u_{\pm}$
have to have the forms
\begin{equation}
u_{+} = \left(\begin{array}{c}
u_{1}\\
\noalign{\vskip 4pt}%
\sigma_{3} u_{1}
\end{array}\right),\quad u_{-} = \left(\begin{array}{c}
u_{2}\\
\noalign{\vskip 4pt}%
-\sigma_{3} u_{2}
\end{array}\right),
\end{equation}
which can be seen from the definitions in
(\ref{representation}). Here $u_{1},u_{2}$ are two component spinors.
Substituting these into (\ref{equation2}), it can be determined that,
for positive energy,
\begin{equation}
u_{+} (\bar{k}) =
\sqrt{\frac{(A-B)\bar{k}_{3}}{2}}\left(\begin{array}{c}
u_{1}\\
\noalign{\vskip 4pt}%
\sigma_{3} u_{1}
\end{array}\right),\quad  u_{-} (\bar{k}) = \frac{1}
{\sqrt{2(A-B)\bar{k}_{3}}}\,\left(\begin{array}{c}
\left(m-\sigma_{i}\bar{k}_{i}\sigma_{3}\right) u_{1}\\
\noalign{\vskip 4pt}%
-\sigma_{3} \left(m-\sigma_{i}\bar{k}_{i}\sigma_{3}\right) u_{1}
\end{array}\right).\label{uspinor}
\end{equation}
The spinors are normalized such that the positive energy spinor $u =
u_{+} + u_{-}$ satisfies the conventional normalization
\begin{equation}
\bar{u} (\bar{k}) u (\bar{k}) = 2m.\label{normalization}
\end{equation}
We note that with the choices
\begin{equation}
u_{1}^{(\uparrow)} = \left(\begin{array}{c}
1\\
\noalign{\vskip 4pt}%
0
\end{array}\right),\quad u_{1}^{(\downarrow)} = \left(\begin{array}{c}
0\\
\noalign{\vskip 4pt}%
1
\end{array}\right),
\end{equation}
we can obtain the spin-up and spin-down spinor states. Thus, for
example, we have
\begin{eqnarray}
u^{(\uparrow)} & = &
\frac{1}{\sqrt{2(A-B)\bar{k}_{3}}}\left(\begin{array}{c}
(A-B)\bar{k}_{3} + m\\
\noalign{\vskip 4pt}%
-(\bar{k}_{1}+i\bar{k}_{2})\\
\noalign{\vskip 4pt}%
(A-B)\bar{k}_{3} - m\\
\noalign{\vskip 4pt}%
-(\bar{k}_{1}+i\bar{k}_{2})
\end{array}\right),\quad
u_{+}^{(\uparrow)} = \sqrt{\frac{(A-B)\bar{k}_{3}}{2}}
\left(\begin{array}{c}
1\\
\noalign{\vskip 4pt}%
0\\
\noalign{\vskip 4pt}%
1\\
\noalign{\vskip 4pt}%
0
\end{array}\right),\nonumber\\
\noalign{\vskip 4pt}%
u^{(\downarrow)} & = &
\frac{1}{\sqrt{2(A-B)\bar{k}_{3}}} \left(\begin{array}{c}
\bar{k}_{1}-i\bar{k}_{2}\\
\noalign{\vskip 4pt}%
(A-B)\bar{k}_{3} + m\\
\noalign{\vskip 4pt}%
-(\bar{k}_{1}-i\bar{k}_{2})\\
\noalign{\vskip 4pt}%
-(A-B)\bar{k}_{3} + m
\end{array}\right),\quad 
\quad u_{+}^{(\downarrow)} =
\sqrt{\frac{(A-B)\bar{k}_{3}}{2}}\left(\begin{array}{r}
0\\
\noalign{\vskip 4pt}%
1\\
\noalign{\vskip 4pt}%
0\\
\noalign{\vskip 4pt}%
-1
\end{array}\right).
\end{eqnarray}

The charge conjugate spinors are obtained from the defining relation
\begin{equation}
v = C \bar{u}^{T} = i \gamma^{2}u^{*},
\end{equation}
which leads, for example, to
\begin{eqnarray}
v^{(\uparrow)} & = &
\frac{1}{\sqrt{2(A-B)\bar{k}_{3}}}\left(\begin{array}{c}
-(\bar{k}_{1}-i\bar{k}_{2})\\
\noalign{\vskip 4pt}%
-(A-B)\bar{k}_{3} + m\\
\noalign{\vskip 4pt}%
\bar{k}_{1}-i\bar{k}_{2}\\
\noalign{\vskip 4pt}%
(A-B)\bar{k}_{3}+m
\end{array}\right),\quad v_{+}^{(\uparrow)} =
\sqrt{\frac{(A-B)\bar{k}_{3}}{2}}\left(\begin{array}{r}
0\\
\noalign{\vskip 4pt}%
-1\\
\noalign{\vskip 4pt}%
0\\
\noalign{\vskip 4pt}%
1
\end{array}\right),\nonumber\\
\noalign{\vskip 4pt}%
v^{(\downarrow)} & = &
\frac{1}{\sqrt{2(A-B)\bar{k}_{3}}}\left(\begin{array}{c}
-(A-B)\bar{k}_{3} + m\\
\noalign{\vskip 4pt}%
\bar{k}_{1}+i\bar{k}_{2}\\
\noalign{\vskip 4pt}%
-(A-B)\bar{k}_{3}-m\\
\noalign{\vskip 4pt}%
\bar{k}_{1}+i\bar{k}_{2}
\end{array}\right),\quad v_{+}^{(\downarrow)} =
\sqrt{\frac{(A-B)\bar{k}_{3}}{2}} \left(\begin{array}{r}
-1\\
\noalign{\vskip 4pt}%
0\\
\noalign{\vskip 4pt}%
-1\\
\noalign{\vskip 4pt}%
0
\end{array}\right).
\end{eqnarray}
It is easy to check that when $B=-A=-1$, these spinors coincide with
the ones in the conventional light-front frame
\cite{harindranath}. From their forms, it is also easily verified that
\begin{equation}
\sum_{s} u_{+, a} (\bar{k},s) u_{+, b}^{\dagger} (\bar{k},s) =
(A-B)\bar{k}_{3} P^{+}_{ab} = \sum_{s} v_{+, a} (\bar{k},s)v_{+,
b}^{\dagger} (\bar{k},s),\label{completeness}
\end{equation}
which will be useful later.

The field decomposition for the independent field component can now be
carried out (much like in the scalar case) as
\begin{equation}
\psi_{+, a} (\bar{x}) = \frac{1}{(2\pi)^{3/2}} \sum_{s} \int
\mathrm{d}^{2}\bar{k}_{\perp} \int_{0}^{\infty}
\frac{\mathrm{d}\bar{k}_{3}}{2\bar{k}_{3}}\left[e^{-i\tilde{\bar{k}}\cdot
\bar{x}}\,b(\bar{k},s)u_{+, a} (\tilde{\bar{k}},s) +
e^{i\tilde{\bar{k}}\cdot 
\bar{x}}\,d^{\dagger} (\bar{k},s) v_{+, a} (\tilde{\bar{k}},s)\right],
\end{equation}
where we have used the definitions in (\ref{ktilde}) and
(\ref{energy}) and note in particular that 
\[
\bar{k}_{0} = \bar{\omega} =
\frac{\omega_{\bar{k}}^{2}}{2|A-B|\bar{k}_{3}} > 0.
\]
This, therefore, truly leads to a separation of the fields into
positive and negative frequency terms which is quite crucial in the
use of the positive and negative energy spinors. 
Requiring the fields to satisfy the anti-commutation relation
(\ref{comm2}), it can be determined with the use of
(\ref{completeness}) that the nontrivial
anti-commutation relation satisfied by the operators $b,d$ has the
form
\begin{equation}
\left[b (\bar{k},s), b^{\dagger} (\bar{k}',s')\right]_{+} = 2 \bar{k}_{3}\,
\delta_{ss'} \delta^{3} (\bar{k}-\bar{k}') = \left[d (\bar{k},s),
d^{\dagger} (\bar{k}', s')\right]_{+}.
\end{equation}
Once again, we see that the basic anti-commutation relations of the
field variables is frame independent and the entire frame dependence
is contained in the plane wave and the spinor solutions.
It follows now from a direct calculation that
\begin{eqnarray}
\langle 0|\psi_{+, a} (\bar{x}) \psi_{+, b}^{\dagger}
(\bar{y})|0\rangle & = & \frac{|A-B| P^{+}_{ab}}{2 (2\pi)^{3}} \int
\mathrm{d}^{2}\bar{k}_{\perp} \int_{0}^{\infty}
\mathrm{d}\bar{k}_{3}\,e^{-i \tilde{\bar{k}}\cdot
(\bar{x}-\bar{y})},\nonumber\\
\noalign{\vskip 4pt}%
\langle 0|\psi_{+, b}^{\dagger} (\bar{y}) \psi_{+, a}
(\bar{x})|0\rangle & = & \frac{|A-B| P^{+}_{ab}}{2 (2\pi)^{3}} \int
\mathrm{d}^{2}\bar{k}_{\perp} \int_{0}^{\infty}
\mathrm{d}\bar{k}_{3}\,e^{i \tilde{\bar{k}}\cdot (\bar{x}-\bar{y})},
\end{eqnarray}
where we have used (\ref{completeness}). The fermion propagator for
the ``$+$'' component can now be determined
to be
\begin{eqnarray}
iS_{F, ab}^{(++)} (\bar{x}-\bar{y}) & = & \langle 0|T\left(\psi_{+, a}
(\bar{x}) \psi_{+, b}^{\dagger} (\bar{y})\right)|0\rangle = \theta
(\bar{x}^{0}-\bar{y}^{0}) \langle 0|\psi_{+, a} (\bar{x}) \psi_{+,
b}^{\dagger} (\bar{y})|0\rangle - \theta (\bar{y}^{0}-\bar{x}^{0})
\langle 0|\psi_{+, b}^{\dagger} (\bar{y}) \psi_{+, a}
(\bar{x})|0\rangle \nonumber\\
\noalign{\vskip 4pt}%
& = & \int
\frac{\mathrm{d}^{4}\bar{k} \left(\sqrt{-\bar{g}}\right)^{-1}}{(2\pi)^{4}}\,
\frac{i(A-B)\bar{k}_{3} P^{+}_{ab}}{\bar{k}^{2} - m^{2} +
i\epsilon}\,e^{-i\bar{k}\cdot (\bar{x}-\bar{y})},
\end{eqnarray}
where we have used the definition of the step function in
(\ref{stepfunction}). 

The other components of the propagator can be obtained now using the
constraint relations. In the momentum space, all the components of the
propagator can be written in the matrix form as
\begin{eqnarray}
\left(\begin{array}{cc}
iS_{F}^{(++)} (\bar{k}) & iS_{F}^{(+-)} (\bar{k})\\
\noalign{\vskip 4pt}%
iS_{F}^{(-+)} (\bar{k}) & iS_{F}^{(--)} (\bar{k})
\end{array}\right)  & = & \left(\begin{array}{cc} 
P^{+} & 0\\
\noalign{\vskip 4pt}%
0 & P^{-}\end{array}\right)\left(\begin{array}{cc}
(A-B)\bar{k}_{3} & \left(-\alpha_{i}\bar{k}_{i} + \frac{m}{A-B}
\bar{\gamma}^{3}\right)\\
\noalign{\vskip 4pt}%
\left(-\alpha_{i}\bar{k}_{i} + \frac{m}{2} \bar{\gamma}^{0}\right) &
\left(2\bar{k}_{0} + (A+B) \bar{k}_{3}\right)
\end{array}\right)\left(\begin{array}{cc}
P^{+} & 0\\
\noalign{\vskip 4pt}%
0 & P^{-}
\end{array}\right)\nonumber\\
\noalign{\vskip 4pt}%
 &  & \qquad\times\, \frac{i}{\bar{k}^{2} - m^{2} + i\epsilon}.
\end{eqnarray}
For $A=0, B=1$, we note that this coincides with the propagators
derived in \cite{das} 
(where the projection operators were not present because of the space
in which the propagators were defined).

\section{Gauge Field Theory}

In this section, we will quantize the Yang-Mills theory with the gauge
fields belonging to $SU(N)$ group in the
conventional light-cone gauge in order to
see if a doubly transverse gauge propagator \cite{brodsky1} results in
the general
frame as well. The action for the theory in the general frame has the form
\begin{equation}
S = \int \mathrm{d}^{4}\bar{x}\,\sqrt{-\bar{g}}\,{\cal L},
\end{equation}
where the Lagrangian density is given by
\begin{eqnarray}
{\cal L} & = &
-\frac{1}{4}\,\bar{g}^{\mu\lambda}\bar{g}^{\nu\rho}\,\bar{F}_{\mu\nu}^{\alpha}
\bar{F}_{\lambda\rho}^{\alpha} \nonumber\\
\noalign{\vskip 4pt}%
 & = & \frac{1}{2}\,(A-B)^{2}
\bar{F}_{03}^{\alpha}\bar{F}_{03}^{\alpha} + (A-B) \bar{F}_{0i}^{\alpha}
\bar{F}_{3i}^{\alpha} + \frac{1}{2}\,
(A^{2}-B^{2})\,\bar{F}_{3i}^{\alpha}\bar{F}_{3i}^{\alpha} -
\frac{1}{4}\,\bar{F}_{ij}^{\alpha} \bar{F}_{ij}^{\alpha}.
\end{eqnarray}
Here, $\alpha=1,2,\cdots , N^{2}-1$ and the field strength tensors is
defined as
\begin{equation}
\bar{F}_{\mu\nu}^{\alpha} = \bar{\partial}_{\mu}
\bar{A}_{\nu}^{\alpha} - \bar{\partial}_{\nu}\bar{A}_{\mu}^{\alpha} +
g f^{\alpha\beta\gamma} \bar{A}_{\mu}^{\beta}\bar{A}_{\nu}^{\gamma},
\end{equation}
with $f^{\alpha\beta\gamma}$ representing the structure constants of
the group. The Abelian limit can be obtained simply by setting the
coupling constant $g$ to zero. We will omit the ``bars'' on the field
variables in the following for simplicity.

The conjugate momenta are now obtained to be
\begin{eqnarray}
\Pi^{0\alpha} & = & \frac{\partial \left(\sqrt{-\bar{g}} {\cal
    L}\right)}{\partial \bar{\partial}_{0}A_{0}^{\alpha}} = 0,\nonumber\\
\noalign{\vskip 4pt}%
\Pi^{3\alpha} & = & \frac{\partial \left(\sqrt{-\bar{g}} {\cal
    L}\right)}{\partial \bar{\partial}_{0}A_{3}^{\alpha}} =
    \sqrt{-\bar{g}}\,(A-B)^{2}\,F_{03}^{\alpha},\nonumber\\
\noalign{\vskip 4pt}%
\Pi^{i\alpha} & = & \frac{\partial \left(\sqrt{-\bar{g}} {\cal
    L}\right)}{\partial \bar{\partial}_{0}A_{i}^{\alpha}} =  {\rm
    sgn}\, (A-B)\, F_{3i}^{\alpha}.
\end{eqnarray}
Thus, we see that the theory has two primary constraints,
\begin{equation}
\chi_{1}^{\alpha} = \Pi^{0\alpha} \approx 0,\quad
\chi_{2i}^{\alpha} = \Pi^{i\alpha} - {\rm sgn}\,
(A-B) F_{3i}^{\alpha} \approx 0.\label{primary3}
\end{equation}
Adding these primary constraints, the starting Hamiltonian density for
the Hamiltonian analysis takes the form
\begin{eqnarray}
{\cal H} & = & \frac{1}{2\sqrt{-\bar{g}}
  (A-B)^{2}}\,\Pi^{3\alpha}\Pi^{3\alpha} - A_{0}^{\alpha}
  \left(\bar{D}_{i}\Pi^{i\alpha} + \bar{D}_{3}\Pi^{3\alpha}\right) -
  \frac{\sqrt{-\bar{g}}
  (A^{2}-B^{2})}{2}\,F_{3i}^{\alpha}F_{3i}^{\alpha} +
  \frac{\sqrt{-\bar{g}}}{4}\, F_{ij}^{\alpha}F_{ij}^{\alpha}\nonumber\\
\noalign{\vskip 4pt}%
 &  & \qquad +\lambda_{1}^{\alpha}\chi_{1}^{\alpha} + \lambda_{2
  i}^{\alpha} \chi_{2i}^{\alpha},
\end{eqnarray} 
where $\lambda_{1}^{\alpha},\lambda_{2 i}^{\alpha}$ are Lagrange
multipliers and the covariant derivative is defined as
\begin{equation}
\bar{D}_{\mu}\Pi^{\nu\alpha} = \bar{\partial}_{\mu}\Pi^{\nu\alpha} + g
f^{\alpha\beta\gamma} A_{\mu}^{\beta}\Pi^{\nu\gamma}.
\end{equation}
In writing
the Hamiltonian density in this form, we have discarded a total
divergence term. We can now use the equal time ($\bar{x}^{0}$)
canonical Poisson brackets
\begin{equation}
\left\{A_{\mu}^{\alpha} (\bar{x}), \Pi^{\nu\beta} (\bar{y})\right\} =
\delta^{\alpha\beta}\delta_{\mu}^{\,\nu}\,\delta^{3} (\bar{x}-\bar{y}),
\end{equation}
with all others vanishing. Requiring
the primary constraints to be stationary, determines the Lagrange
multiplier $\lambda_{2i}^{\alpha}$ and leads to the secondary constraint
\begin{equation}
\chi_{3}^{\alpha} = \bar{D}_{i}\Pi^{i\alpha} + \bar{D}_{3}\Pi^{3\alpha}
\approx 0.\label{secondary3}
\end{equation}
Furthermore, requiring the secondary constraint (\ref{secondary3}) to
be stationary, determines the Lagrange multiplier
$\lambda_{1}^{\alpha}$ and the chain of constraints terminates.

The complete set of constraints for the theory are given by
(\ref{primary3}) and (\ref{secondary3}). It can be easily checked that
of these $\chi_{1}^{\alpha},\chi_{3}^{\alpha}$ correspond to first
class constraints
while $\chi_{2i}^{\alpha}$ represents a second class constraint. This is
consistent with the general characteristics of a light-front theory,
namely, it develops a genuine second class constraint in addition to
the ones already present in the conventional theory. Since there are
two first class constraints, we choose two gauge fixing conditions
that will make these second class. Keeping the physical light-cone gauge in
mind, we choose the gauge fixing conditions to correspond to
\begin{equation}
\phi_{1}^{\alpha} = \bar{n}\cdot A^{\alpha} \approx 0,\quad
\phi_{2}^{\alpha} =
\bar{\partial}\cdot A^{\alpha} \approx 0,\quad \bar{n}^{\mu} =
(0,0,0,1),\quad \bar{n}^{2}=0.\label{gaugefixing}
\end{equation}
The Dirac brackets can now be determined iteratively and we simply
note the final result for the equal time Dirac brackets involving the
field variables,
\begin{equation}
\left\{A_{\mu}^{\alpha} (\bar{x}),A_{\nu}^{\beta}
    (\bar{y})\right\}_{D} =  \frac{\delta^{\alpha\beta}\,{\rm
    sgn}\, (A-B)}{2}\,P_{\mu\nu}^{T}
    (\bar{n},\bar{\partial})\,\bar{\partial}_{3}^{-1}\,\delta^{3}
    (\bar{x}-\bar{y}),
\end{equation}
where we have defined (all the derivatives are with respect to the
argument $\bar{x}$)
\begin{equation}
\bar{\partial}_{3}^{-1}\,\delta^{3} (\bar{x}-\bar{y}) = \delta^{2}
(\bar{x}_{\perp}-\bar{y}_{\perp}) \epsilon(\bar{x}^{3}-\bar{y}^{3}),
\end{equation}
and $P_{\mu\nu}^{T} (\bar{n},\bar{\partial})$ represents the
projection operator transverse to both
$\bar{n}_{\mu},\bar{\partial}_{\mu}$ and has the form
\begin{equation}
P_{\mu\nu}^{T} (\bar{n},\bar{\partial}) = \bar{g}_{\mu\nu} -
\frac{\bar{n}_{\mu}\bar{\partial}_{\nu}+\bar{n}_{\nu}\bar{\partial}_{\mu}}
{\bar{n}\cdot  \bar{\partial}} +
\frac{\bar{\partial}^{2}}{(\bar{n}\cdot
  \bar{\partial})^{2}}\,\bar{n}_{\mu}\bar{n}_{\nu}.\label{projection3}
\end{equation} 
The projection operator is symmetric, homogeneous in both
$\bar{n}_{\mu},\bar{\partial}_{\mu}$ and it is straightforward to
check that it satisfies
\begin{equation}
\bar{n}^{\mu} P_{\mu\nu}^{T} (\bar{n},\bar{\partial}) = 0 =
\bar{\partial}^{\mu} P_{\mu\nu}^{T} (\bar{n},\bar{\partial}).
\end{equation}
The other equal time brackets can be obtained from this since the
constraints can be strongly set equal to zero in the Dirac
brackets. This also leads to the fact that in the quantum theory, we
can take the independent commutation relation involving fields to be
\begin{equation}
\left[A_{\mu}^{\alpha} (\bar{x}),A_{\nu}^{\beta} (\bar{y})\right] =
    \frac{i\delta^{\alpha\beta}\,{\rm
    sgn}\, (A-B)}{2}\,P_{\mu\nu}^{T}
    (\bar{n},\bar{\partial})\,\delta^{2}
    (\bar{x}_{\perp}-\bar{y}_{\perp}) \epsilon
    (\bar{x}^{3}-\bar{y}^{3}).\label{comm3}
\end{equation}

The field decomposition can now be carried out in the standard manner
as in the case of the scalar field and takes the form
\begin{equation}
A_{\mu}^{\alpha} (\bar{x}) = \frac{1}{(2\pi)^{3/2}} \sum_{h=1}^{2} \int
\mathrm{d}^{2}\bar{k}_{\perp} \int_{0}^{\infty}
\frac{\mathrm{d}\bar{k}_{3}}{2\bar{k}_{3}}\,\epsilon_{\mu}
(\bar{k},h)\left[e^{-i\tilde{\bar{k}}\cdot \bar{x}}\,a^{\alpha} (\bar{k},h) +
e^{i\tilde{\bar{k}}\cdot \bar{x}}\,a^{\dagger\,\alpha}
(\bar{k},h)\right],\label{field3} 
\end{equation}
where, as before, we have identified
\begin{equation}
\bar{k}_{0} = \bar{\omega} =
\frac{\omega_{\bar{k}}^{2}}{2|A-B|\bar{k}_{3}} >0,
\end{equation}
and have used (\ref{ktilde}). The two physical polarization vectors
are chosen to be real for simplicity and are supposed to satisfy
\begin{equation}
\bar{n}\cdot \epsilon (\bar{k},h) = 0 = \bar{k}\cdot \epsilon
(\bar{k},h),\quad \epsilon(\bar{k},h)\cdot \epsilon (\bar{k},h') =
-\delta_{hh'},\quad  h=1,2.\label{polarization}
\end{equation}
We can choose, for example,
\begin{equation}
\epsilon_{\mu} (\bar{k},1) =
\frac{1}{(A-B)\bar{k}_{3}}\,(\bar{k}_{1}, (A-B)\bar{k}_{3}, 0,
0),\quad \epsilon_{\mu} (\bar{k},2) =
\frac{1}{(A-B)\bar{k}_{3}}\,(\bar{k}_{2}, 0, (A-B)\bar{k}_{3},
0),\label{polarization1} 
\end{equation}
which satisfy the properties in (\ref{polarization}). In addition to
these two physical polarization vectors, we can also choose two other
vectors 
\begin{equation}
\epsilon_{\mu} (\bar{k},3) = \bar{k}_{\mu},\quad \epsilon_{\mu}
(\bar{k},4) = \bar{n}_{\mu},
\end{equation}
such that together they define a set of basis vectors. From the
completeness of these vectors, it is easy to show that
\begin{equation}
\sum_{h=1}^{2} \frac{\epsilon_{\mu} (\bar{k},h)\epsilon_{\nu}
  (\bar{k},h)}{\epsilon (\bar{k},h)\cdot \epsilon (\bar{k},h)} +
  \frac{\epsilon_{\mu} (\bar{k},3)\epsilon_{\nu} (\bar{k},4) +
  \epsilon_{\nu} (\bar{k},3)\epsilon_{\mu} (\bar{k},4)}{(\epsilon
  (\bar{k},3)\cdot \epsilon (\bar{k},4))} - \frac{\epsilon
  (\bar{k},3)\cdot \epsilon (\bar{k},3)}{(\epsilon (\bar{k},3)\cdot
  \epsilon (\bar{k},4))^{2}}\,\epsilon_{\mu} (\bar{k},4)
  \epsilon_{\nu} (\bar{k},4) = \bar{g}_{\mu\nu},
\end{equation}
which leads to
\begin{equation}
\sum_{h=1}^{2} \epsilon_{\mu} (\bar{k},h) \epsilon_{\nu} (\bar{k},h) =
- P_{\mu\nu}^{T} (\bar{n},\bar{k}) = - \bar{g}_{\mu\nu} +
\frac{\bar{n}_{\mu}\bar{k}_{\nu}+\bar{n}_{\nu}\bar{k}_{\mu}}{(\bar{n}\cdot
\bar{k})} - \frac{\bar{k}^{2}}{(\bar{n}\cdot
\bar{k})^{2}}\,\bar{n}_{\mu}\bar{n}_{\nu}.\label{projection4}
\end{equation}
This can also be constructed directly from the outer product of the forms of
the polarization vectors given in (\ref{polarization1}).

Requiring the fields (\ref{field3}) to satisfy the commutation
relations in (\ref{comm3}), we can determine that the nontrivial
commutation relation involving the operators
$a^{\alpha},a^{\dagger\,\alpha}$ has the
form
\begin{equation}
\left[a^{\alpha} (\bar{k},h), a^{\dagger\,\beta} (\bar{k}',h')\right] =
2\,\bar{k}_{3}\,\delta^{\alpha\beta}\,\delta_{hh'}\,\delta^{3}
(\bar{k}-\bar{k}').
\end{equation}
The Feynman propagator for the theory is now straightforward to
calculate. We note that
\begin{equation}
\langle 0|A_{\mu}^{\alpha} (\bar{x}) A_{\nu}^{\beta}
(\bar{y})|0\rangle = - \frac{\delta^{\alpha\beta}}{(2\pi)^{3}} \int
\mathrm{d}^{2}\bar{k}_{\perp} 
\int_{0}^{\infty}
\frac{\mathrm{d}\bar{k}_{3}}{2\bar{k}_{3}}\,P_{\mu\nu}^{T} (\bar{n},
\bar{k})\,e^{-i \tilde{\bar{k}}\cdot (\bar{x}-\bar{y})}.
\end{equation}
With this we obtain,
\begin{eqnarray}
iG_{F, \mu\nu}^{\alpha\beta} (\bar{x}-\bar{y}) & = & \langle 0|
T\left(A_{\mu}^{\alpha} 
(\bar{x}) A_{\nu}^{\beta} (\bar{y})\right)|0\rangle = \theta
(\bar{x}^{0}-\bar{y}^{0}) \langle 0|A_{\mu}^{\alpha} (\bar{x})
A_{\nu}^{\beta} 
(\bar{y})|0\rangle + \theta (\bar{y}^{0}-\bar{x}^{0}) \langle
0|A_{\nu}^{\beta} (\bar{y}) A_{\mu}^{\alpha} (\bar{x})|0\rangle \nonumber\\
\noalign{\vskip 4pt}%
 & = & \int
\frac{\mathrm{d}^{4}\bar{k}
  \left(\sqrt{-\bar{g}}\right)^{-1}}{(2\pi)^{4}}\,\frac{-i
\delta^{\alpha\beta}P_{\mu\nu}^{T}
  (\bar{n},\bar{k})} {\bar{k}^{2} + i\epsilon}\,e^{-i\bar{k}\cdot
  (\bar{x}-\bar{y})}.
\end{eqnarray}
It follows, therefore, that in momentum space, the Feynman propagator
takes the form
\begin{equation}
iG_{F, \mu\nu}^{\alpha\beta} (\bar{k}) = - \left(\bar{g}_{\mu\nu} -
\frac{\bar{n}_{\mu}\bar{k}_{\nu}+\bar{n}_{\nu}\bar{k}_{\mu}}{(\bar{n}\cdot
  \bar{k})} +\frac{\bar{k}^{2}}{(\bar{n}\cdot
  \bar{k})^{2}}\,\bar{n}_{\mu}\bar{n}_{\nu}\right)
\frac{i\delta^{\alpha\beta}}{\bar{k}^{2}+i\epsilon}.\label{prop}
\end{equation}
We note that for any value of $A,B$, this propagator is doubly
transverse, namely,
\begin{equation}
\bar{n}^{\mu} G_{F, \mu\nu}^{\alpha\beta} (\bar{k}) = 0 =
  \bar{k}^{\mu} G_{F, \mu\nu}^{\alpha\beta} (\bar{k}).
\end{equation}
This has been observed earlier \cite{brodsky1} in the conventional
light-front 
quantization. However, our analysis shows that this is a generic
feature in the general light-front frame and is a consequence of the
particular choice of gauge fixing in (\ref{gaugefixing}).

We note here that the propagator (\ref{prop}) has poles at
$\bar{n}\cdot \bar{k} = 0$ in addition to the usual pole at
$\bar{k}^{2}=0$. This necessitates a prescription for handling such
poles which in the conventional equal time theories is given by the
Leibbrandt-Mandelstam prescription \cite{leibbrandt}
\begin{equation}
\frac{1}{n\cdot k} \rightarrow \lim_{\epsilon\rightarrow
0^{+}}\,\frac{1}{n\cdot k + i\epsilon (\tilde{n}\cdot k)},
\end{equation}
where $\tilde{n}^{\mu}$ is a dual light-like vector with $n\cdot
\tilde{n} \neq 0$. Since the Leibbrandt-Mandelstam prescription
involves only scalar combinations and as we have already argued scalar
quantities do not change under a change of frame, this prescription
can be readily extended to the general light-front frame. For our
choice of the light-like vector in (\ref{gaugefixing}), we have 
\begin{equation}
\tilde{\bar{n}}^{\mu} = \left(2,0,0, A+B\right).
\end{equation}
In our analysis, we have chosen a physical gauge fixing condition
since we are interested in the Hamiltonian quantization of the theory
with physical degrees of freedom. However, other gauge fixing
conditions may be more useful from the path integral point of view
that we have not pursued here. Some of these have been discussed in
\cite{das} within the path integral approach. 

\section{Conclusion}

In this paper we have studied systematically the quantization of
quantum field theories in a general light-front frame. We have carried
out the Hamiltonian analysis for scalar, fermion as well as 
gauge theories. The decomposition of the fields into positive and
negative frequency parts is done carefully which leads to frame
independent (anti) commutation relations for the annihilation and
creation operators. In the case of scalar fields and the gauge fields,
the frame dependence is contained completely in the plane wave
functions (as well as in the polarization vectors), while in the case of
the Dirac field, the spinor solutions are frame
dependent as well and we have derived these explicitly. The
propagators for the various fields have been obtained from the vacuum
expectation values of the time ordered products and they coincide, for
specific values of $A,B$, with the earlier known results. In
particular, we have shown that in the light-cone gauge, the gauge
propagator is doubly transverse in any general frame much like it was
observed to be in the conventional light-front frame \cite{brodsky1}. 

We conclude by saying that all of our analysis has been carried
out with a view to constructing a
thermal Hilbert space within the formalism of thermofield dynamics
which will then allow us to study various operatorial questions,
discussed in the introduction, at finite temperature. We hope to be
able to report on such a finite temperature analysis in a future
publication. 

\vskip 1cm

\noindent{\bf Acknowledgment:}
\medskip

We would like to thank Profs. J. Frenkel and S. Okubo for helpful
comments. This work was
supported in part by US DOE Grant number DE-FG 02-91ER40685 and by
CNPq, Brazil.


\begin{thebibliography}{10}

\bibitem{dirac} P. A. M. Dirac, Rev. Mod. Phys. {\bf 21}, 392 (1949);
{\em Lectures in Quantum Mechanics} (Benjamin, New York, 1964);
A. J. Hanson, T. Regge and C. Teitelboim, {\em Constrained Hamiltonian
Systems} (Academia Nazionale dei Lincei, Rome, 1976).

\bibitem{fubini} S. Fubini and G. Furlan, Physics {\bf 1}, 229 (1965);
S. Weinberg, Phys. Rev. {\bf 150}, 1313 (1966); L. Susskind,
Phys. Rev. {\bf 165}, 1535 (1968); K. Bardakci and M. B. Halpern,
Phys. Rev. {\bf 176}, 1686 (1968).

\bibitem{kogut} J. B. Kogut and D. E. Soper, Phys. Rev. {\bf D1}, 2901
(1970).  

\bibitem{chang} S-J. Chang, R. G. Root and T. M. Yan, Phys. Rev. {\bf
D7}, 1133 (1973); S-J. Chang and T. M. Yan, {\em ibid.} {\bf D7}, 1147
(1973).  

\bibitem{brodsky0} P. P. Srivastava and S. J. Brodsky, Phys. Rev. {\bf
  D61}, 025013 (2000).

\bibitem{brodsky1} P. P. Srivastava and S. J. Brodsky, Phys. Rev. {\bf
D64}, 045006 (2001).

\bibitem{brodsky} There are numerous papers on the subject for us to
list. Instead, we are going to refer the readers to only a few well
written review articles which can serve as sources for other
references. M. Burkhadt, Adv. Nucl. Phys. {\bf 23}, 1 (1996);
S. J. Brodsky, H. C. Pauli and S. S. Pinsky, Phys. Rep. {\bf 301}, 299
(1998); K. Yamawaki, ``Zero mode problem on the light-front'',
hep-th/9802037; T. Heinzl, ``Light-cone quantization: Foundations and
applications'', Lec. Notes Phys. {\bf 572}, 55 (2001).

\bibitem{das0} V. S. Alves, A. Das and S. Perez, Phys. Rev. {\bf D66},
125008 (2002); A. Das, hep-th/0310247, Proceedings of {\em Hadrons and Beyond
(LC 03)}, Durham, England, 2003.

\bibitem{weldon} H. A. Weldon, Phys. Rev. {\bf D67}, 085027 (2003). 

\bibitem{das} A. Das and X. Zhou, Phys. Rev. {\bf D68}, 065017
(2003).
 
\bibitem{gomis} A general light-front frame has been considered
earlier from a quite different perspective by E. Elizalde and
J. Gomis, Nuc. Phys. {\bf B122}, 535 (1977).

\bibitem{umezawa} H. Umezawa, H. Matsumoto and M. Tachiki, {\em
Thermofield Dynamics and Condensed States} (North-Holland, Amsterdam,
1982). 

\bibitem{das1} A. Das, {\em Finite Temperature Field Theory} (World
Scientific, Singapore, 1997).

\bibitem{harindranath} A. Harindranath, hep-ph/9612244, in {\em
Light-front Quantization and Non-perturbative QCD}, J.P. Vary and
F. W\"{o}lz (eds.), International Institute of Theoretical and Applied
Physics, ISU, Ames, USA, ISBN: 1-891815-00-8.

\bibitem{bjorken} B. J. Bjorken and S. Drell, {\em Relativistic
  Quantum Mechanics} (McGraw-Hill, New York, 1965).

\bibitem{srivastava} P. P. Srivastava, {\em Perspectives of
  Light-front Quantized Field Theory}, in {\em Quantum Field Theory},
  A. Mitra (ed.), Hindustan Book Agency, New Delhi, 2000.

\bibitem{leibbrandt} G. Leibbrandt, Rev. Mod. Phys. {\bf 59}, 1067
(1987). 


\end{thebibliography}
\end{document}